\documentclass[conference]{IEEEtran}
\IEEEoverridecommandlockouts
\usepackage{url}
\usepackage{dsfont}

\usepackage{cite}
\usepackage{amsmath,amssymb,amsfonts}
\usepackage{algorithmic}
\usepackage{graphicx}
\usepackage{textcomp}
\usepackage{xcolor}
\def\BibTeX{{\rm B\kern-.05em{\sc i\kern-.025em b}\kern-.08em
    T\kern-.1667em\lower.7ex\hbox{E}\kern-.125emX}}
\begin{document}

\title{Advancing Gene Selection in Oncology: A Fusion of Deep Learning and Sparsity for Precision Gene Selection\\
}

\author{\IEEEauthorblockN{Akhila Krishna}
\IEEEauthorblockA{\textit{Department of Electrical Engineering} \\
\textit{IIT Bombay}\\
Mumbai, India \\
19d070006@iitb.ac.in
}
\and
\IEEEauthorblockN{Ravi Kant Gupta}
\IEEEauthorblockA{\textit{Department of Electrical Engineering} \\
\textit{IIT Bombay}\\
Mumbai, India \\
184070025@iitb.ac.in
}
\and
\IEEEauthorblockN{Pranav Jeevan}
\IEEEauthorblockA{\textit{Department of Electrical Engineering} \\
\textit{IIT Bombay}\\
Mumbai, India \\
194070025@iitb.ac.in
}
\and
\IEEEauthorblockN{Amit Sethi}
\IEEEauthorblockA{\textit{Department of Electrical Engineering} \\
\textit{IIT Bombay}\\
Mumbai, India \\
asethi@iitb.ac.in
}

}

\maketitle

\begin{abstract}
Gene selection plays a pivotal role in oncology research for improving outcome prediction accuracy and facilitating cost-effective genomic profiling for cancer patients. This paper introduces two gene selection strategies for deep learning-based survival prediction models. The first strategy uses a sparsity-inducing method while the second one uses importance based gene selection for identifying relevant genes. Our overall approach leverages the power of deep learning to model complex biological data structures, while sparsity-inducing methods ensure the selection process focuses on the most informative genes, minimizing noise and redundancy. Through comprehensive experimentation on diverse genomic and survival datasets, we demonstrate that our strategy not only identifies gene signatures with high predictive power for survival outcomes but can also streamlines the process for low-cost genomic profiling. The implications of this research are profound as it offers a scalable and effective tool for advancing personalized medicine and targeted cancer therapies. By pushing the boundaries of gene selection methodologies, our work contributes significantly to the ongoing efforts in cancer genomics, promising improved diagnostic and prognostic capabilities in clinical settings.
\end{abstract}


\section{Introduction}

Cancer stands as a prominent global cause of mortality, responsible for almost 10 million deaths in 2020, equating to nearly one in six recorded deaths. Early detection and effective treatment can result in the successful cure of numerous cancers. 
To facilitate the precise selection of a treatment plan, predicting the risk of patient mortality in a particular time becomes essential in ensuring optimal therapeutic decision-making
~\cite{who}.

Effective treatment planning can be done based on models that can accurately predict patient survival probability. While various data, including patient history, medical imaging, and genomic data, offer means for such prognostication, our focus lies specifically on utilizing genomic data. Our model does the analysis based on whole transcriptomic RNA sequencing data from their tumour biopsy or tumor resection~\cite{kumar2023learning}. However, the challenge arises from the high dimensionality of genomic data and the limited survival patient information, leading to the curse of dimensionality~\cite{chattopadhyay2019gene} in machine learning models. To address this, we advocate for gene selection as a vital dimension reduction strategy. In contrast to traditional dimension reduction methods, such as principal component analysis (PCA) ~\cite{mackiewicz1993principal}, our approach emphasizes gene selection, offering a solution not only to the curse of dimensionality but also contributing to essential research on identifying cancer-specific gene signatures. An additional benefit of gene selection is reduction in the cost of genomic testing by reducing the size of the gene panel.

Survival datasets often include censored observations where the event of interest has not occurred by the end of the study~\cite{fotso2018deep}. In response to this problem, we introduce two distinct gene selection methods using survival models which can model censored data: 1) Gene selection using a modified neural multi-task logistic regression (NMTLR), which incorporating L1 regularization loss~\cite{shalev2009stochastic} to the original NMTLR model ~\cite{fotso2018deep}, and 2) gene importance learning with survival risk prediction, which is a novel strategy to discard the unimportant genes. The former identifies common genes across various cancers, while the latter focuses on cancer-specific gene selection.  By introducing these methodologies, our study aims to advance the understanding of genomic data-based survival prediction and contribute valuable insights to cancer research.

\section{Related Work}
The fast-developing 'omics technologies, such as DNA micro-array that can help assess expression levels across thousands of genes, pose significant analytical challenges due to the number of variables these can extract. The high-dimensional and low sample size datasets generated by these technologies necessitate sophisticated tools for extracting meaningful biological insights~\cite{pavithra2017feature,bihani2014comparative}. 
Clustering, a pivotal unsupervised technique, emerges as a versatile solution across fields like bio informatics~\cite{iqbal2020orienting} and image analysis~\cite{han2001spatial}, enhancing gene data analysis. The evolution of feature selection, especially through innovative methods like SCAD~\cite{frigui2000simultaneous} and KBCGS~\cite{chen2016kernel}, underscores the shift towards integrating clustering for refined attribute discrimination. Concurrently, deep learning, particularly auto encoders, has revolutionized unsupervised learning, offering nuanced data transformation and clustering~\cite{song2014deep,chen2019hybrid}. Evolutionary algorithms enhance this landscape by offering robust strategies for feature reduction ~\cite{salem2017classification,ghosh2019genetic,jansi2019two,tiwari2017approach}. This marks a paradigm shift in addressing high-dimensional, label-scarce datasets with unparalleled precision and efficiency.

Gene selection methodologies in machine learning are categorized into supervised, unsupervised, and semi-supervised approaches. Supervised gene selection~\cite{filippone2006supervised} leverages task knowledge, but faces challenges such as over-fitting and data mislabeling~\cite{ang2015supervised}. 
Unsupervised selection~\cite{filippone2005unsupervised} operates on unlabeled data, and offers unbiased insights, but it fails to account for gene interactions, which can compromise discriminative power~\cite{acharya2017unsupervised}. Semi-supervised selection~\cite{sheikhpour2017survey} blends labeled and unlabeled data to enhance class separation and uncover the feature space's geometry. Hence, it represents a balanced approach that mitigates the limitations of purely supervised or unsupervised methods. 

In survival analysis~\cite{kalbfleisch2011statistical}, the Cox proportional hazards model~\cite{cox1972regression} and other parametric survival distributions prediction methods (multi-task logistic regression model (MTLR) and NMTLR) have long been used to fit the survival time of a population. Several enhancements to MTLR have been proposed, including NMF-MTLR ~\cite{kumar2023learning} and Bayesian neural network-based MTLR in~\cite{loya2020uncertainty} and~\cite{loya2019bayesian} , aiming to enhance survival prediction capabilities. However, these methods primarily concentrate on enhancing survival prediction rather than visualizing the genes accountable for such improvements.

\begin{figure*} 
\centering
\includegraphics[height=6cm,width=17.5cm]{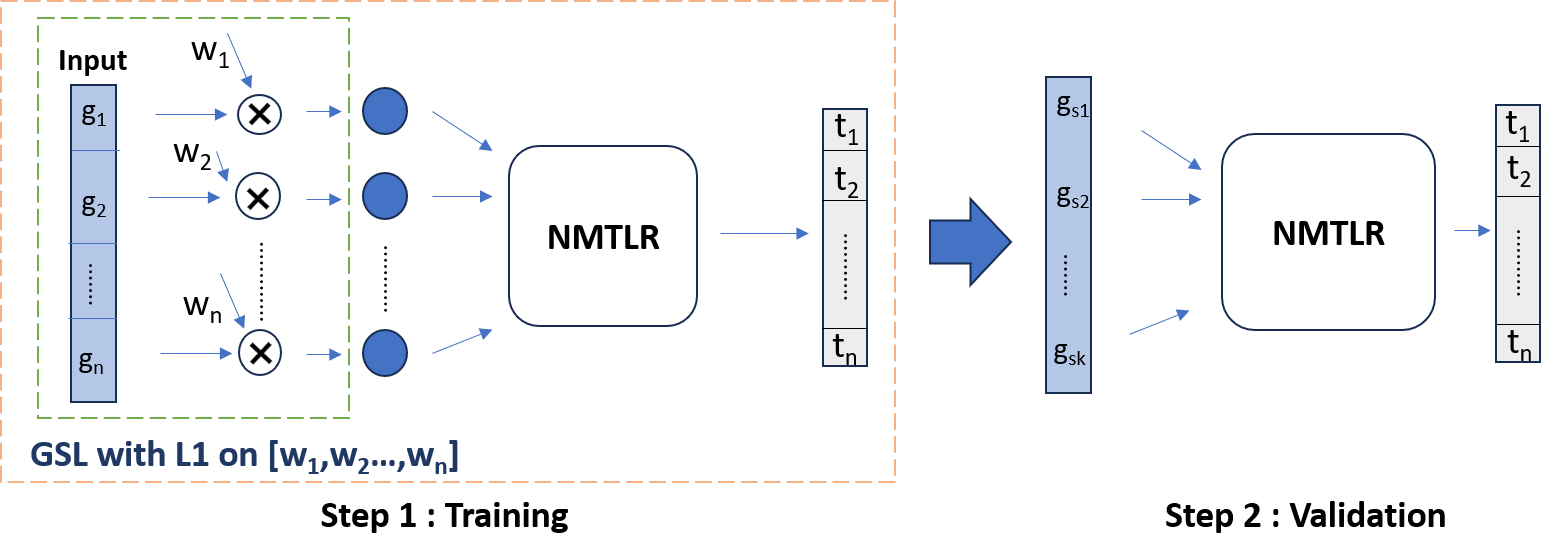}
\caption{Overall model framework for sparse-NMTLR model : 1) Step 1 - $[g_1,g_2...g_n]$ is the input data, which is then multiplied with $[w_1,w_2...w_n]$, which is the weight denoting the importance of each gene and L1 regularization is applied on these weights.
2) Step 2 - The k selected genes ($g_{s1},g_{s_2}...g_{s_k}$) from Step 1 is used to train another NMTLR model.} 
\label{fig1}
\end{figure*}

\section{Materials and Method}
\subsection{Dataset}
The dataset includes information from 10,156 patients obtained from the TCGA portal. These patients span 33 TCGA projects (primary cancer sites) and involve clinical, gene expression, and treatment details for 33 tumor types under the TCGA-PanCancer dataset ~\cite{liu2018integrated} .

The following steps were undertaken to preprocess the dataset:

1. Log Transformation of Gene Expression Data:
Raw gene expression values were transformed using $log (1 + x)$ to maintain numerical stability in subsequent analyses.

2. Data Normalization:
To ensure smooth training, gene expression and clinical data, having different variances, were normalized to a unit Gaussian distribution ~\cite{dutka1991fundamentals}.

3. Gene Selection:
A focused gene selection process identified and retained 746 genes based on criteria available on the Sanger website ~\cite{cancer}.

4. One-Hot Encoding of Clinical Features
\begin{figure*} 
\centering
\includegraphics[height=4.5cm,width=17.5cm]{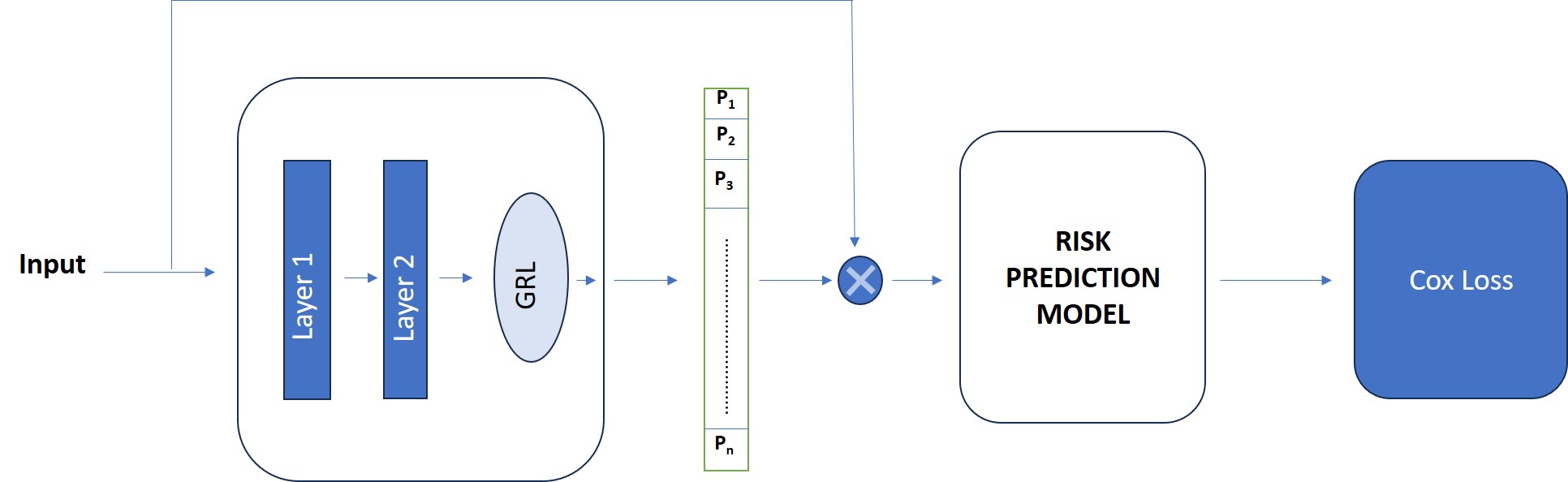}
\caption{Overall model framework for GIL Network : The GRL network gives the $[P_1, P_2, .... P_n]$ which gives the probability of contribution of each gene to the survival of patients. It is then multiplied with input to get the overall effect of each gene which is then passed to risk prediction model.}
\label{main}
\end{figure*}
\subsection{Sparsity induced deep learning model}
A Gene Selection Learning (GSL) approach in \ref{fig1} was formulated, leveraging L1 regularization to induce sparsity. This framework was integrated with a neural multi-task logistic regression model (NMTLR)~\cite{fotso2018deep}. The weights produced by the gene selection framework were used to determine the significance of each gene. Following this, L1 regularization was applied to these weights, enabling the elimination of genes identified as unimportant and enhancing the precision of the gene selection process. The analysis algorithm is devised in 2 stages : (1) Training Phase - The model is trained with L1 regularization applied on the GSL layer weights. The genes with non-zeros weights is selected (2) Validation - The selected genes from first phase is used for training the NMTLR model and is then evaluated on test data. This approach ensures a robust evaluation of the selected genes.

\subsection{Patient-specific Gene Importance Learning Deep Learning Network (GIL)}

In the model, shown in \ref{main}, we introduced a gene importance learning network to filter out less crucial genes using the Gene Removal Layer (GRL), which is defined as 
: \\
$$ GRL = ReLU(Softmax(v)-\alpha)$$ 
where $v$ is the input to GRL which is the quantity signifying the importance of each gene and $\alpha$ is a hyper parameter which is used to remove least important genes. The GRL, combined with linear layers, forms the gene importance learning framework. The output from this framework is multiplied with the input, and is fed into the Deep Learning-based survival risk prediction model. To assess the model, we utilized the Efron's partial likelihood~\cite{efron1977efficiency} which take ties and censoring into account (Cox loss function)~\cite{lin1989robust}. The Cox loss function is formulated as follows: 
\begin{multline}
\sum_{j} \sum_{i \in H_j}\log \left( F(x_i) \right) -  \\ \sum_{j}\sum_{s=0}^{m_j-1} \log \left( \sum_{i: t_j\leq T_i} F(x_i) -  \frac{s}{m_j} \sum_{i \in H_j}\log(F(x_i) ) \right)
\end{multline}
where $T_i$ = time of censoring or time of death for $i^{th}$ patient (the event is denoted by $E_i=0$ for alive patients and $E_i = 1$ for dead patients), $H_j =  \{ i; T_i = t_j$ and $E_i =1 \}$ (which denotes the risk set at time $t_j$), $ m_j = |H_j|$, $F(x_i) = exp(f(x_i))$ and $f(x_i)$ is the non-linear function for risk prediction.


\subsection{Experimentation}
\subsubsection{Sparse Model}
In the experiments conducted, the Adam optimizer~\cite{kingma2014adam} with a learning rate of $5\times 10^{-4}$ was employed for the training. Since sparse model was not giving patient specific genes, we did the analysis on 3 cancers (BRCA, OV and HNSC) which were having the highest number of data. A Sparse Neural Multi Task Logistic Regression (sparse-NMTLR) model was trained on this data to identify the top 164 genes based on the weights assigned to the initial layer of the model. The L1 regularization parameter was set at 10, while the L2 regularization parameter was set at 0.01 and the model is run for 500 epochs until the C-index was falling drastically.  The genes with the non-zero weights at the initial layer of the model (164 genes) were taken alone and used to train the model again and then tested on our test data to validate our gene selection algorithm. The results are mentioned in Table \ref{tab1}.

\subsubsection{Gene Importance Learning Algorithm}
In the course of our experimentation, we employed the Adam optimizer with a learning rate set at $10^{-5}$, executing the code over 400 epochs until convergence was achieved. The focus of our analysis centered on identifying the most common genes within the subset of the most powerful genes for test data patients across various cancers. In this context, "most common" denotes genes present in over 50\% of patients, while "most powerful" refers to the top 25\% of genes based on their respective gene importance values. Detailed insights into this analysis can be found in Table \ref{tab2}.
\\
\\

The dataset was divided in the ratio 75:12.5:12.5 for training, validation, and testing for both of the methods. The metric used for evaluation is the concordance index (C-index). 
 $$ C-index =\frac{\sum_{i,j}\mathds{1}_{T_j<T_i}.\mathds{1}_{r_j>r_i}.\delta_j}{\sum_{i,j}\mathds{1}_{T_j<T_i}.\delta_j }$$
 where $r_i$ is the predicted risk of the $i^{th}$ patient.

\section{Results and Discussion}

We introduce a new approach for learning the importance of genes using the softmax function. This method outperforms the Cox model when applied on the TCGA dataset. It has surpassed Cox model in the individual C-index of all cancers except kidney clear cell carcinoma (KIRC) and colon adenocarcinoma (COAD). It showed a substantial improvement, surpassing the Cox model by almost 0.2 in the overall C-index, indicating its effectiveness in enhancing survival prediction accuracy.

Our study with sparse-NMTLR model focused on BRCA, OV, and HNSC cancers, leading to the identification of a set of 164 genes that significantly contribute to the survival of these cancers, as confirmed by our validation framework. The gene subset was analysed and the top 10 genes are the following : POLE, WAS, BRCA1, PTK6, CHST11, EZH1, AKAP9, TEC, HMGA1, ATP2B3, NCOA4 and PTPRB. The whole list with the gene weights is available in ~\cite{sparse_mtlr_genes}. The gene subset includes many that have been proven to be relevant for these cancers in the previous studies ~\cite{rosen2003brca1}, ~\cite{chen2023prognostic}, ~\cite{herman2015chst11} . Comparing different models, we found that the Cox model performed better than NMTLR and MTLR models.

Examining the role of clinical features in survival, our GIL-based model outperformed traditional frameworks. Additionally, the inclusion of clinical features resulted in a notable increase in the C-index. Interestingly, when we focused on a selected subset of genes for analysis, we observed an additional boost in the C-index. This suggests that by avoiding the challenges associated with high-dimensional data, a targeted gene subset can improve predictive performance. 

The future work involves creating a gene selection algorithm that takes into account the co-expression mechanism of genes. Additionally, we will explore the possibility of integrating images with genomic data to enhance overall performance.

\begin{table}[htbp]
\caption{Results of sparse-NMTLR model}
\begin{center}
\begin{tabular}{|c|c|c|c|c|}
\hline

\textbf{Method} & \textbf{BRCA}& \textbf{HNSC} & \textbf{OV}  & \textbf{Overall}\\
\hline
Cox (with all genes) & 0.66 & 0.67 & 0.83 & 0.80\\
\hline
Cox (with 164 genes) & 0.73 & \textbf{0.68} & \textbf{0.85} & \textbf{0.81} \\
\hline
MTLR (with all genes) & 0.66 & 0.63 & 0.83 & 0.80 \\
\hline
 MTLR (with 164 genes) & 0.73 & 0.65 & 0.84 & \textbf{0.81} \\
 \hline 
NMTLR (with all genes) & 0.61 & 0.60 & 0.81 & \textbf{0.81} \\
\hline
NMTLR (with 164 genes) & \textbf{0.76} & 0.63 & 0.84  & \textbf{0.81} \\
\hline 
\end{tabular}
\label{tab1}
\end{center}
\end{table}


\begin{table}[htbp]
\caption{Results of GIL network}
\begin{center}
\begin{tabular}{|c|c|c|c|}
\hline

\textbf{Cancer type} & \textbf{Cox}& \textbf{GIL}& \textbf{GIL with clinical}\\
\hline
BRCA & 0.64 & 0.67$\pm$0.003 & \textbf{0.69$\pm$0.004} \\
\hline
UCEC & 0.62 & 0.72$\pm$ 0.004 & \textbf{0.74$\pm$0.003}\\
\hline
KIRC & \textbf{0.73} & 0.72$\pm$ 0.010 & 0.69 $\pm$ 0.015\\
\hline 
HNSC & 0.59 & \textbf{0.60$\pm$0.020} & 0.58 $\pm$ 0.017\\ 
\hline 
LGG & 0.78  & \textbf{0.88 $\pm$0.014} & 0.85 $\pm$ 0.013 \\ 
\hline 
SKCM & 0.62 & \textbf{0.63$\pm$ 0.013}  & 0.57 $\pm$ 0.020\\
\hline 
LUAD & 0.60 & 0.62$\pm$ 0.005 & \textbf{0.62 $\pm$0.002}\\ 
\hline 
COAD & \textbf{0.64} & 0.60 $\pm$ 0.025 & 0.52 $\pm$0.063 \\
\hline 
BLCA & 0.57 & \textbf{0.58$\pm$ 0.005} & 0.56 $\pm$0.007 \\
\hline
Total & 0.73 & 0.75$\pm$0.005 & \textbf{0.76$\pm$ 0.002} \\
\hline
\end{tabular}
\label{tab2}
\end{center}
\end{table}





\bibliographystyle{named}
\bibliography{EMBS_MTLR}

\end{document}